\documentclass[MSNbibl,nameyear,dvips]{arxstspdf}
\usepackage{flushend}
\usepackage{stfloats}
\usepackage{graphicx}
\usepackage{url,breakurl}

\volume{29}
\issue{1}
\pubyear{2014}
\firstpage{128}
\lastpage{143}
\doi{10.1214/13-STS437} 

\makeatletter

\newproclaim{example}{Example}

\def\cal{\mathcal}
\newcommand{\eqref}[1]{(\ref{#1})}
\makeatother

\begin{document}
\begin{frontmatter}

\title{A Parametric Framework for the Comparison of Methods of Very
Robust Regression}
\runtitle{A Parametric Framework for the Comparison of Methods of Very
Robust Regression}

\begin{aug}
\author[a]{\fnms{Marco} \snm{Riani}\corref{}\ead[label=e1]{mriani@unipr.it}\ead[label=u1,url]{http://www.riani.it}},
\author[b]{\fnms{Anthony C.} \snm{Atkinson}\ead[label=e2]{a.c.atkinson@lse.ac.uk}}
\and
\author[c]{\fnms{Domenico} \snm{Perrotta}\ead[label=e3]{domenico.perrotta@ec.europa.eu}}
\runauthor{M. Riani, A.~C. Atkinson and D. Perrotta}

\affiliation{Universit\`a di Parma, London School of Economics and
European Commission Joint Research Centre}

\address[a]{Marco Riani is Professor,
Dipartimento di Economia, Universit\`a di Parma, Via Kennedy 6, 43100
Parma, Italy \printead{e1,u1}.}
\address[b]{Anthony Atkinson is Emeritus Professor, Department of
Statistics, London School of Economics,
London WC2A 2AE, United Kingdom \printead{e2}.}
\address[c]{Domenico Perrotta is Senior Research Scientist, European
Commission Joint Research Centre, Via E. Fermi 2749, I-21027 Ispra (VA),
Italy \printead{e3}.}
\end{aug}

%
\begin{abstract}
There are several methods for obtaining very robust estimates of
regression parameters that asymptotically resist 50\% of outliers in
the data. Differences in the behaviour of these algorithms depend on
the distance between the regression data and the outliers. We introduce
a parameter $\lambda$ that defines a parametric path in the space of
models and enables us to study, in a systematic way, the properties of
estimators as the groups of data move from being far apart to close
together. We examine, as a function of $\lambda$, the variance and
squared bias of five estimators and we also consider their power when
used in the detection of outliers. This systematic approach provides
tools for gaining knowledge and better understanding of the properties
of robust estimators.
\end{abstract}

%

\end{frontmatter}

\section{Introduction}

Multiple regression is one of the main tools of applied statistics. It
has, however, long been appreciated that ordinary least squares as a
method of fitting regression models is exceptionally susceptible to the
presence of outliers.
Instead, very robust methods, that asymptotically resist 50\% of
outliers, are to be preferred. Our paper presents a systematic,
parameterised framework for the nonasymptotic comparison of these methods.

Very robust regression was introduced by Rous\-seeuw (\citeyear{rou:84}) who developed
suggestions of
\citet{hamp:75} that led to the Least Median of Squares (LMS) and Least
Trimmed Squares (LTS)
algorithms. For some history of more recent developments see \citet
{rous+vand:2006}. More general discussions of robust methods are in
\citet{yohai:2006} and \citet{mor:07}. We illustrate our methods for
the comparison of high-breakdown regression procedures with comparisons
of the performance of LTS and other well-established methods, including
S and MM estimators, with that of a publicly available algorithm for
very robust regression that uses the Forward Search (FS). See
\citet{arc:2010} for a recent discussion of the FS.

Very robust regression estimators share the property that,
asymptotically, they have a breakdown point of 50\% (see Section~\ref{breakdown}) as the main data and outliers become infinitely far apart.
In order to distinguish between the estimators we study, in a
systematic way, their properties as the distance between the two groups
of observations decreases. In Section~\ref{ours} we introduce a
parameterised framework, with parameter $\lambda$, for moving the
outliers along a trajectory which is initially remote from the main
data, but which then passes close to it before again becoming far away.
We control whether, at their closest, the two populations share the
same centre. We design measures of overlap to calibrate the trajectories.

Numerical results are in Sections~\ref{olapsec} and \ref{pointsec}. In
Section~\ref{olapsec} we take the outliers from the regression model to
have a multivariate normal distribution. This provides a very general
scenario for outliers that can range from a seemingly random scatter
around the regression plane to points virtually on a line. The special
case of point contamination is explored in Section~\ref{pointsec}.
Boxplots of the estimates from the five methods as $\lambda$ varies
indeed show that, for wide separations, the methods have similar
properties. However, they differ markedly as the two populations
converge. In order to summarise this information, we look at cumulative
plots, over the range of $\lambda$, of the variance and squared bias of
the estimators. Another method of comparing robust estimators is by
their properties for outlier detection (\cite{ch:90}). In Section~\ref{olapsec} we calculate power curves as a function of $\lambda$ for the
number of outliers detected. Since the curves indicate that the
estimators provide tests of varying sizes, we find the size of the
outlier tests in Section~\ref{powersec}.

There are two main conclusions. The first is that the parameterised
family of departures provides a cogent framework for investigating the
behaviour of very robust estimators. The second is that we can clearly
establish the properties of the various methods of very robust
regression in terms of the bias and variance of estimators and the size
and power of outlier tests.

The approach is motivated in the next section by an example in which
there is a mixture of two regression lines. Such data arise in the
analysis of trade where different countries or suppliers may report
different relationships between value and quantity. Although, in our
example, there are only two countries, which makes the data appropriate
for a robust analysis assuming one model describes at least half the
data, there is no reason why there should not be several suppliers. The
comparative robust analysis of the data is in Section~\ref{tdsec2}.

\section{An Example: Trade Data}
\label{tdsec}

\begin{figure}

\includegraphics{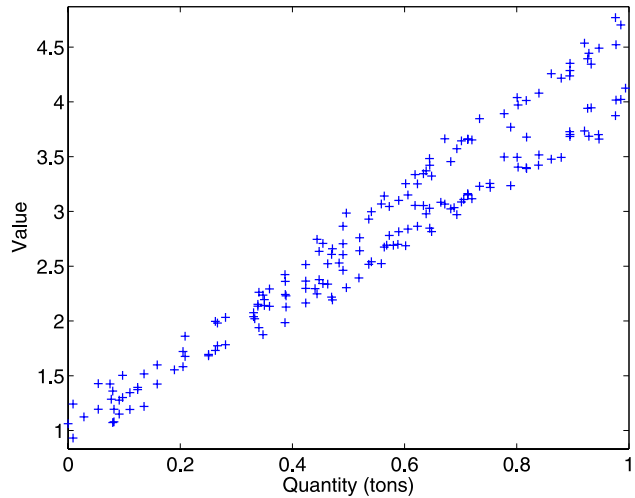}

\caption{Trade data: A mixture of two regression lines.}\label{TDsca}
\end{figure}

Our interest in the behaviour of robust regression procedures when the
main data and outliers are close together was stimulated by a seemingly
simple example with a
single explanatory variable. The data, shown in Figure~\ref{TDsca}, are
of a kind discussed by \citet{dom+al:2009} in the
detection of fraud in international trade, where false declarations of
price are used in tax
evasion and money laundering. The result is data which are a mixture of
regression lines.

There are 180 observations in Figure~\ref{TDsca} that come from two
firms. The structure is of two
lines that overlap for lower values; any kind of separation is likely
to be impossible. However, the
two lines are clearly separate for the higher values of $y$ and $x$ and
a robust procedure should
respond to this pattern, by downweighting some of the observations in
estimation and flagging them
as outliers. If the outlier pattern suggests that there is a mixture of
regression models, the analysis can move to clusters of regression
lines, as in \citet{garcia+:2010}. But the first stage is the
identification of outliers, for which a robust fit is required.
Interest in the analysis is not in individual outliers
but whether the two lines differ. Accordingly, in Section~\ref{ours} we
introduce a Bonferroni
adjustment to provide, at least theoretically, the desired samplewise
size of the
outlier test. We return to the analysis of these data in
Section~\ref{tdsec2}.\looseness=1

\section{Models, Data, Robustness and Methods}
\label{modsec}

\subsection{Outliers and Regression}
\label{robreg}

We consider the usual regression model with random carriers
[\citet{hub+elvz:2009}, page~197]. The observations are i.i.d.
random vectors ($y$, $x^T) \in\Re^{p+1}$, where $y \in\Re$ and $x
\in \Re^p$ satisfy
%
%
\begin{equation}
\label{regrmod} y = x^T \beta+ u.
\end{equation}
The random errors $u$ are distributed independently of the covariates
$x$ and $\beta$ is the $p \times1$ vector parameter of interest.

In the absence of outliers the least squares estimate $\hat{\beta}$ is
the best linear unbiased estimator of $\beta$. However, even a single
outlier can cause $\hat{\beta}$ to be severely biased. Figure~2 of
\citet{rou:84} is a paradigmatic example in which a cluster of 20
outliers at a remote point in   $X$-space cause the least squares
fitted line to pass close to the cluster. The robust line, in that case
LMS, completely downweights the outliers and is close to the least
squares line for the 30 remaining data points when the outliers have
been deleted.

Of course, the outliers are not usually known and the problems of
robust estimation and outlier detection are closely related. In robust
estimation a fit is found which is close to that without the outliers.
The robust fit then allows identification of all important outliers.
However, the outliers may be difficult to identify from a nonrobust fit
since their inclusion can seriously bias the parameter estimates and
make the outliers seem less remote. ``Backward'' methods of outlier
detection that start from a fit to all data and then proceed by
eliminating observations that appear to be outlying can therefore fail.

One example for regression is a synthetic data set due to \citet
{hbk:84} with $n = 75$ and three explanatory variables. The figures on
page~95 of \citet{rl:87} show that the least squares residuals, unlike
those from LMS, are not sufficiently large to call attention to the ten
outlying observations. Numerous other examples for regression are in
Chapters~3 and 4 of \citet{a+r:2000}; further plots for the
\citet{hbk:84} data are on pages~72 and~73.

\subsection{Maxbias and Breakdown Point}
\label{breakdown}

Robustness is concerned with fitting a single model to data which are
generated by two, or maybe
more, models. We suppose that the larger part of the data, $1 - \varepsilon$, where $ 0 < \varepsilon<
0.5$, is generated by the model $M_1(\theta_1)$ and the remaining part
$\varepsilon$ of the data is
generated by the model $M_2(\theta_2)$. In the absence of outliers,
that is, when $\varepsilon=0$, an ideal robust estimator would have a
variance that achieved the Cramer--Rao lower bound. If the data were
contaminated, the estimate would be unbiased. Such estimators do not
exist. \citeauthor{yohai:2006} [(\citeyear{yohai:2006}), Section~3.4] describe some compromises
between the two properties.

Robust methods study the properties of methods that fit
$M_1(\theta_1)$ in ignorance of knowledge of the form of the outlier
generating model
$M_2(\theta_2)$, which can be quite general. When $M_1(\cdot)$ is a
regression model, $M_2(\cdot)$ is often taken, for example, to distribute
observations randomly
over a large space, concentrate them in a cluster or to be a
second regression model. There is no difficulty in having $M_1(\theta)
= M_2(\theta)$, but then we
must have $\theta_1 \ne\theta_2$.

With $M_1(\theta_1)$ the usual regression model \eqref{regrmod}, let $\mathrm{E}
u^2 = \sigma^2 < \infty$ and $I(x) = \mathrm{E} xx^T$. The bias of an
estimator $\hat{\beta}$ of $\beta$ is
%
%
\begin{equation}
b(\hat{\beta}) = \bigl\{(\hat{\beta} - \beta)^TI(x) (\hat{\beta} -
\beta)\bigr\}^{0.5}. \label{bias}
\end{equation}
The bias depends on the estimator, the distribution of $y$ and $x$, the
amount of contamination $\varepsilon$ and on $M_2(\theta_2)$. In the
robustness literature dependence on $M_2(\theta_2)$ is removed by
considering estimators that minimise the maximum (asymptotic) bias
within a particular class of estimators. The maxbias curve shows how
the maximum bias varies with $\varepsilon$. The breakdown point of an
estimator is the minimum value of $\varepsilon$ for which $b(\hat{\beta})$
in \eqref{bias} equals $\infty$. The estimators we consider all have an
asymptotic breakdown point of 50\%. An introduction to these ideas for
regression is given by \citeauthor{yohai:2006} [(\citeyear{yohai:2006}), Section~5.9]. Although
50\% is customarily considered to be the maximum possible breakdown
value, higher values may occur in clustering.

Unfortunately maxbias curves are only calculable for some estimators
and distributions of $x$. The latter are often assumed to be
elliptically symmetrical. A~summary of the literature is given by \citet
{berren+:2001} who extend results on maxbias curves to regression
models with intercepts and to regressors that have Student's $t$ and
Cauchy distributions, although without intercepts. \citet{berren+:2007}
find maxbias curves for MM estimators, again without intercepts. We
describe MM estimators in Section~\ref{fivesec}.

The theoretical results that are available are asymptotic and do not
cover all estimators or models of interest to us. A few numerical
results are available for finite samples. Figure~5.14 of \citet
{yohai:2006} plots biases of several estimators as a function of a
single parameter, the slope of the regression line for the
contaminating observations. Figure~3 of \citet{garcia+:2010} is
more in the spirit of our numerical approach. It shows the simulated
bias as a point cluster of outliers moves around a regression line.
When the outliers are very close to the line, the bias is negligible,
as it is when the outliers are far away and are easily downweighted by,
in this case, LTS. Only for intermediate outliers is the bias appreciable.

In our numerical comparisons we study the variance as well as the bias
of the estimators. In addition, following the comments in Section~\ref{robreg} about the relationship between robustness and outlier
detection, we asses the power of outlier tests using residuals from
robustly fitted models.

\subsection{Five Methods for Very Robust Regression}
\label{fivesec}

We compare and contrast the properties of what are currently considered
the five best methods for very robust regression. The algorithms that
we use are all publicly available from the Forward Search
Data Analysis (FSDA) Matlab toolbox. See \citet{tool:2012}. In this
section we outline the methods that we compare. Full implementation
details of the algorithms are in the documentation of the FSDA library.
Numerically, all algorithms involve selecting many subsets from the
data. An important factor in our ability to conduct as many simulations
as were necessary is the efficient sampling of subsets provided in FSDA
as described by \citet{bench:2012}.

Traditional robust estimators attempt to limit the influence of
outliers by replacing the squares of the residuals in least squares
estimation of $\beta$ by a function $\rho$ of the residuals which is bounded.
Of the numerous forms that have been suggested for $\rho(\cdot)$
(\cite{PRS:72}; \cite{hamp:86}; \cite{hub+elvz:2009}),
we use the most popular choice, Tukey's
Biweight, in which extreme residuals are replaced by the value
$c^2/6$. See, for example, \citet{rl:87}, (4.31). The M-estimator of
scale $\tilde\sigma_M $ is the solution to a second equation, for
example, \citet{rl:87}, (4.30), depending on a second $\rho$ function
and a constant $K_c$. Although the two $\rho$ functions may be
different, we again use the biweight. The minimum value of $\tilde
\sigma_M$ which satisfies this second equation provides the S-estimate
of scale ($\tilde\sigma_S$) with associated estimate of the vector of
regression coefficients ($\tilde\beta_S$). $K_c$ and $c$ are related
constants which are linked to the breakdown point of the estimator of
$\beta$. Fixing the breakdown point at 50\% gives a value for 1.547 for
$c$ and an efficiency for estimation of 28.7\% [\citet{rl:87}, pages~135--143].

The MM-regression estimator is intended to improve the S estimator. The
S estimate of scale $\tilde\sigma_S$ is used and kept fixed to
estimate $\beta$, but with a value of $K_c$ giving a higher efficiency.
Because of the relationship between $K_c$ and $c$, the hope expressed
by \citeauthor{rl:87} [(\citeyear{rl:87}), page~143] is that the MM estimator maintains its high
breakdown point for finite samples. Following the recommendation of
\citeauthor{yohai:2006} [(\citeyear{yohai:2006}), page~126], we take $K_c$ such that the (asymptotic)
nominal efficiency is 85\%, which gave a high-breakdown estimator in
our examples, which included up to 23\% of outliers. Small numerical
experiments indicate that even slight increases, for example, to a
nominal efficiency of 87\%, result in very low breakdown and estimates
similar to those from least squares.

The remaining three estimators of $\beta$ result from more direct
approaches. The forward search (FS) uses least squares to fit subsets
of observations of increasing size $m$ to the data, with $p \le m \le
n$. The forward search for regression was introduced by \citet
{a+r:2000}. A recent general review of forward search methods is
\citet{arc:2010}. For efficient parameter estimation $m$ should
increase until all $n-m$ observations not in the subset used for
fitting are outliers. The outliers
are found by testing at each step
of the search. The effect of simultaneous testing can be severe (\cite
{a+r:2006}); the FS algorithm is designed to have size $\alpha$ of
declaring an outlier free sample to contain at least one outlier. We
perform the outlier test for individual observations at a Bonferronised
size $\alpha^* = \alpha/n$, so taking the $1 -\alpha^*$ cutoff value of
the reference
distribution. In our calculations $\alpha= 0.01$. The automatic
algorithm is based on that of \citet{rac:2009} who used scaled
Mahalanobis distances to detect outliers in multivariate normal data.
For regression we replace these distances by deletion residuals.

{\spaceskip=0.2em plus 0.05em minus 0.05em
In Least Trimmed Squares (LTS)}
 [\citet{rou:84}, page~876] the search
is over subsets of size $h$ for which the residual sum of squares from
least squares estimates of $\beta$ is minimised. LTS has an asymptotic
breakdown point of 50\% when $h = [n/2]+\break  [(p+1)/2]$.

To increase efficiency, reweighted versions of LTS estimators can be
computed. These
reweighted estimators, denoted LTSr, are computed by giving weight 0 to
outlying observations. We then obtain a sample of reduced size $n-k$, possibly
outlier free, to which OLS is applied. For comparison of results from
LTSr with those from the FS, we perform the outlier test at the
Bonferronised size $\alpha^*$.

In FS, LTS and its reweighted version LTSr, $\sigma^2$ is estimated
from subsets formed by hard $(0,1)$ trimming. Consistency factors for the
estimators are given by \citet{croux+r:1992}, equation~(6.5) and follow
from the results of \citet{tall:63} on elliptically trimmed
multivariate normal distributions. For LTS we also use the small sample
correction of \citet{pison+va+will:2002}.

\section{A Parameterised Family of Departures}
\label{ours}

As $y_{M2} \sim M_2(\theta_2) \rightarrow\infty$ the observations
$y_{M1}$ and $y_{M2}$ from the two models become
increasingly well separated. Under these conditions the five estimators
in our study have similar properties. We are also interested in those
data configurations when the
observations are not so well separated, so that both $y_{M1}$ and
$y_{M2}$ may be used in estimating
$\theta$ because of overlap between the two samples. Such
configurations are highly informative about the differences in
properties of robust estimators. We define a finite-sample measure of
the overlap of $y_{M1}$ and $y_{M2}$ that is designed to be informative
for regression models. In general, the properties of robust estimators
depend on the ``distance'' between the two models. Table~3.1 of \citet
{yohai:2006} is a typical example showing the behaviour of robust
estimators as one observation~$\rightarrow\infty$. Our proposed
distance measure likewise provides a framework for comparisons in the
more complicated world of regression procedures.

There is a sample ${\cal S}_1$ of $n_1$ observations from $M_1(\theta
_1)$ with distribution
$F_1(y_i;x_i,\theta_1)$ conditional on the value of $x_i$. These values
of $x_i$ belong to a design region ${\cal X}$. The sample ${\cal S}_2$
of $n_2$ observations from $M_2(\theta_2)$ has conditional
expectation $ \mathrm{ E}(y;x_i,\theta_2)$. Some values of $x_i$ from
${\cal S}_2$ may belong to ${\cal X}$. We define the indicator
%
%
\begin{equation}
I_{i,\gamma} = \cases{ %
= 1,& $\mbox{if }
F_1^{-1}(\gamma/2;x_i,\theta _1)$\vspace*{2pt}\cr
& \hspace*{20pt}$<
\mathrm{ E}(y;x_i,\theta_2)$\vspace*{2pt}\cr
&\hspace*{20pt}$ < F_1^{-1}(1-
\gamma/2;x_i,\theta_1),$\vspace*{2pt}\cr
&\hspace*{20pt}$ i \in{\cal S}_2,
x_i \in {\cal X}$,
\vspace*{2pt}\cr
=  0,& $\mbox{otherwise.}$}
\end{equation}
The index is a function of both $\theta_1$ and $\theta_2$ and we
examine it over a set of parameter values $\Theta_1$ and $\Theta_2$.
For a particular set of parameter values $\theta_{1,k}$ and $\theta
_{2,k}$ the overlapping index is defined as
%
%
\begin{equation}
O_{\gamma,k} = \sum_{i} I_{i,\gamma,k},\quad i
\in{\cal S}_2. \label{defoverlap}
\end{equation}
With $M_1(\theta_1)$ normal theory regression, we are therefore
counting the total number of observations in ${\cal S}_2$ for which
$x_i \in{\cal X}$, the conditional medians of which lie in a strip
around the expectation of $M_1(\cdot)$. As $\gamma$ decreases, the strip
becomes broader in $y$. If also for all $i \in{\cal S}_2, x_i \in
{\cal X}$, then ${ O}_{\gamma,k} \rightarrow n_2$, the number of
observations in ${\cal S}_2$.

It is informative to keep $\theta_1$ fixed and to vary $\theta_2$ in a
smooth way with a parameter $\lambda\in\mathfrak{R}$. Then we look at
a set of indexes
%
%
\begin{equation}
{\cal O}_{\gamma}(\lambda) = \{O_{\gamma,k}\},\quad \theta _{1}
\in\Theta_1 \mbox{ and } \theta_{2,k} \in
\Theta_2(\lambda). \label{defoverlap2}
\end{equation}
In particular, we vary $\theta_2$ linearly using the combination
%
%
\begin{eqnarray} \label{lambdathet}
\theta_{2,k} = \lambda_k \theta_2^0
+ (1 - \lambda_k)\theta_2^1
\nonumber
\\[-8pt]
\\[-8pt]
 \eqntext{(-\infty<
\lambda_k \in\Lambda< \infty).}
\end{eqnarray}
The set $\Lambda$ of values considered is problem dependent. With
$\theta_2^0 = \theta_1$ the centre of $M_2$ passes through that of
$M_1$. Other choices of $\theta_2^0$ can produce a trajectory in which
the observations $y_2$ are always outlying. Our examples show how the
variance and bias of the parameter estimates change in a smooth way
with $\lambda$, but in different and informative ways for different estimators.

\begin{figure*}

\includegraphics{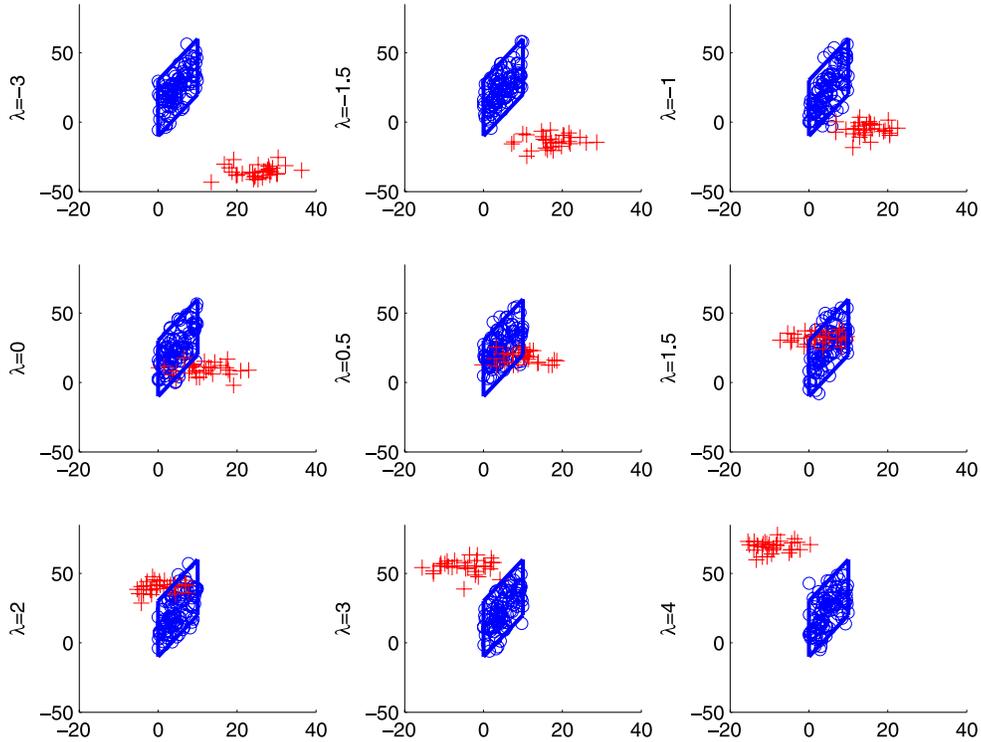}

\caption{Example \protect\ref{ex1}. Typical simulated data sets with $n_1 =100$ and
$n_2 = 30$
for nine values of $\lambda$. As $\lambda$ increases, observations from
$M_2$ become close to those from $M_1$ and then become remote again.
The parallelogram defines the region for the empirical overlapping
index.}\label{Sim1sca}
\end{figure*}

In Section~\ref{olapsec} the contamination $M_2$ in our examples comes from a multivariate
normal distribution. In the \hyperref[appA]{Appendix} we show how to calculate the
probability of intersection between this distribution and a strip
around the regression plane. We call this the theoretical overlapping
index. Although it ignores ${\cal X}$, it does signal cases where $y_2$
lies close to the regression line, even if remote from~${\cal X}$.
These observations would then be ``good'' leverage points, in the sense
that they improve the estimates of the regression parameters. For
counting vertical outliers we need observations that lie in ${\cal X}$.
These are signalled by the index defined in \eqref{defoverlap}, which
has to be calculated by simulation. We therefore call this the
empirical index.

\section{The Numerical Effect of Overlap: Normal Contamination}
\label{olapsec}

Because of the flexibility of our systematic approach, we can
potentially cover a wide range of possibilities. Here we look at three
numerical examples with normal contamination. In the next section we
consider point contamination. We look at boxplots of the estimates over
a suitable $\Lambda$ and relate these plots to the overlapping indices.
We separate out the variance and bias components of the estimates and
compare these through cumulative plots over $\Lambda$. Finally, we
compare the estimators for their power of detecting outlying
observations, that is, those that come from model 2. The detection of
outliers is particularly important if we require an indication that other
methods of data analysis are appropriate.

In our one-variable regression examples $M_1$ is the regression model
$y_i = \alpha+ \beta x_i + \varepsilon_i$, with the independent $x_i \sim
U(a,b)$, these values generated once for all observations and values of
$\lambda$. The standard deviation of $Y$ is $\sigma_{\varepsilon}$ and overlapping
indices were calculated for a strip of width $\pm2\sigma_{\varepsilon}$
around $\mathrm{E}(Y)$.

%
\begin{figure*}

\includegraphics{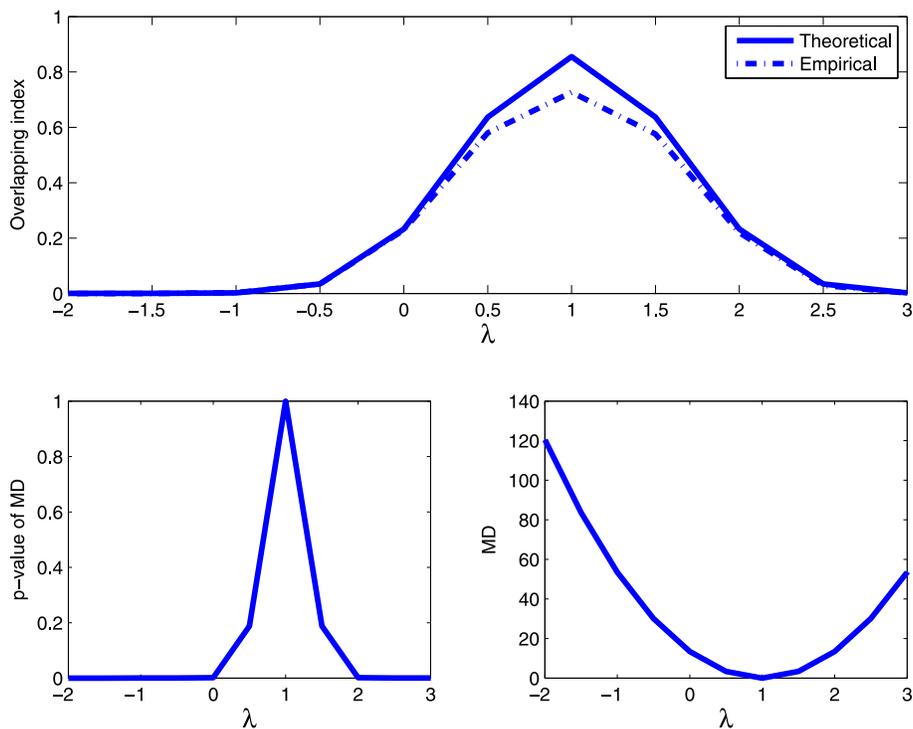}

\caption{Example \protect\ref{ex1}. Upper panel: theoretical and empirical overlapping
indices for the data in Figure~\protect\ref{Sim1sca}, showing maxima at
$\lambda= 1$. Lower panel: squared Mahalanobis distance of $M_1$ from
$M_2$ (right) and corresponding $p$-values (left).} \label{Sim1overlap}
\end{figure*}

The expectation of $x$ is $\mu_{x} =(a+b)/2$. The bivariate normal
distribution for $M_2$ has mean $\mu$ and variance $\Sigma$ given by
%
%
\begin{eqnarray}\label{bvM2}
\quad\mu&= &\pmatrix{ \alpha+ \beta(\mu_{x}
+ d)
\vspace*{2pt}\cr
\mu_{x} + d}
 \lambda + \pmatrix{
\mu_2
\vspace*{2pt}\cr
\mu_2} (1- \lambda) \quad\mbox{and}
\nonumber
\\[-8pt]
\\[-8pt]
\nonumber
 \Sigma&=& \pmatrix{\sigma_1^2 & \sigma_{12}
\vspace*{2pt}\cr
 \sigma_{12} & \sigma_2^2},
\end{eqnarray}
where the first component corresponds to the response. When $\lambda=
1$ the centres of the two populations are identical when the
displacement $d = 0$.

\begin{example}\label{ex1}
We took $n_1 = 100$ with $\alpha= 10, \beta= 3,
\sigma_{\varepsilon} = 10, a = 0$ and $b = 10$. For the second population,
$n_2 = 30, \sigma_1^2 = \sigma_2^2=20, \sigma_{12} = 2$ and $\mu_2 = 10$. Also, $d = 0$
so the centres coincide at $\lambda= 1$. There were 100 simulations
for each value of $\lambda$.
\end{example}

Figure~\ref{Sim1sca} shows nine typical simulated data sets. As $\lambda
$ increases from $-3$ to 4, the centre of $M_2$ passes through that of
$M_1$, at which point there is almost complete overlapping of the
observations from the two populations. That the overlap is not complete
is shown by the plots of the indices in the upper panel of Figure~\ref{Sim1overlap},
the maxima of which are less than one. The theoretical index is
slightly higher than the empirical index, as there is some probability
of observations falling within the band of $y$ values that are not in
${\cal X}$. On the other hand, the plot of the squared Mahalanobis
distance from the mean of $M_2$ to that of $M_1$ has a minimum of zero,
showing identity of the two centres.

\begin{figure*}

\includegraphics{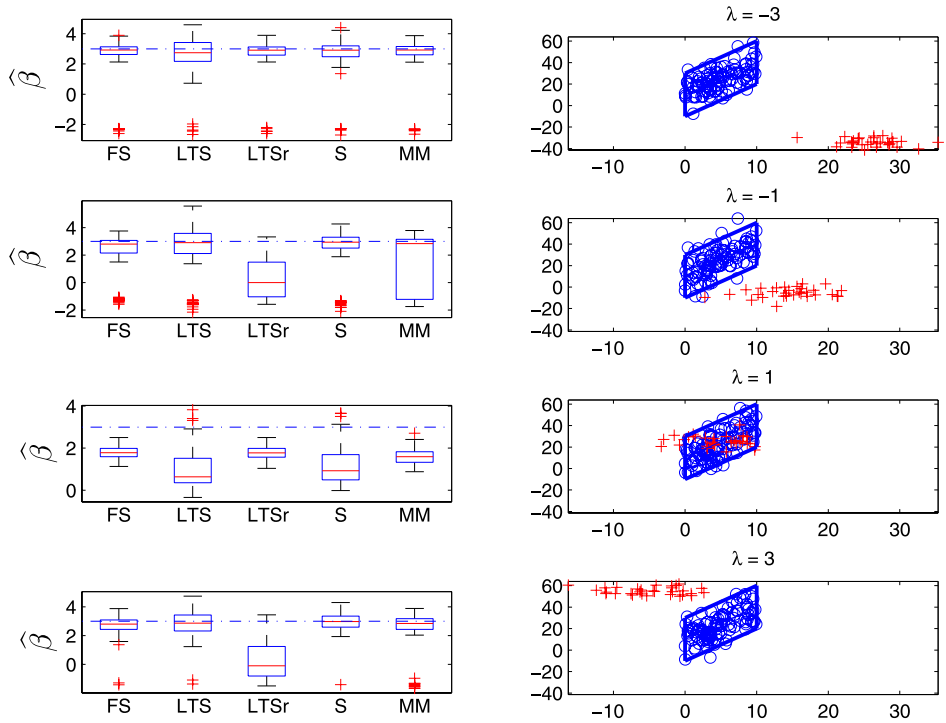}

\caption{Example \protect\ref{ex1}. Four simulated data sets for $ \lambda = -3, -1,
1$ and 3. Left-hand panels: boxplots, from 100 simulations, of
estimates of $\beta$ (dotted and dashed line: $\beta_1=3$) for FS, LTS, LTSr S and
MM estimators. Right-hand panels: typical
simulations for these four values of $\lambda$.}\label{Sim1boxsca}
\end{figure*}

We now consider the effect of these data configurations on the
estimation of $\beta$. The left-hand panels of Figure~\ref{Sim1boxsca}
show boxplots, from 100 simulations, of the values of
the five estimators for a series of values of $\lambda$, together with
a typical data configuration for each. For $\lambda= -3$, observations
from $M_2$ lie below and to the right of those from $M_1$. If these
outliers are not identified, the slope of the line is decreased. The
boxplots all show some simulations where such estimates occur. LTS has
the highest variance amongst the estimators in the main part of the
boxplot, that is, for the estimates when all outlying observations are
rejected, with S the second most variable. For $\lambda= -1 $, LTSr
and MM are most affected by the outliers. The value $\lambda= 1$
corresponds to virtually complete overlap of the two groups. All
methods, on average, give estimates that are biased downwards. However,
those for LTS and S are both more variable and more biased. In the last
panel, for $\lambda= 3$, the outliers are not as well separated as
they are in panel 1. LTSr now has appreciable negative bias, due to the
inclusion of outliers in the reweighting stage.\looseness=-1

\begin{figure*}

\includegraphics{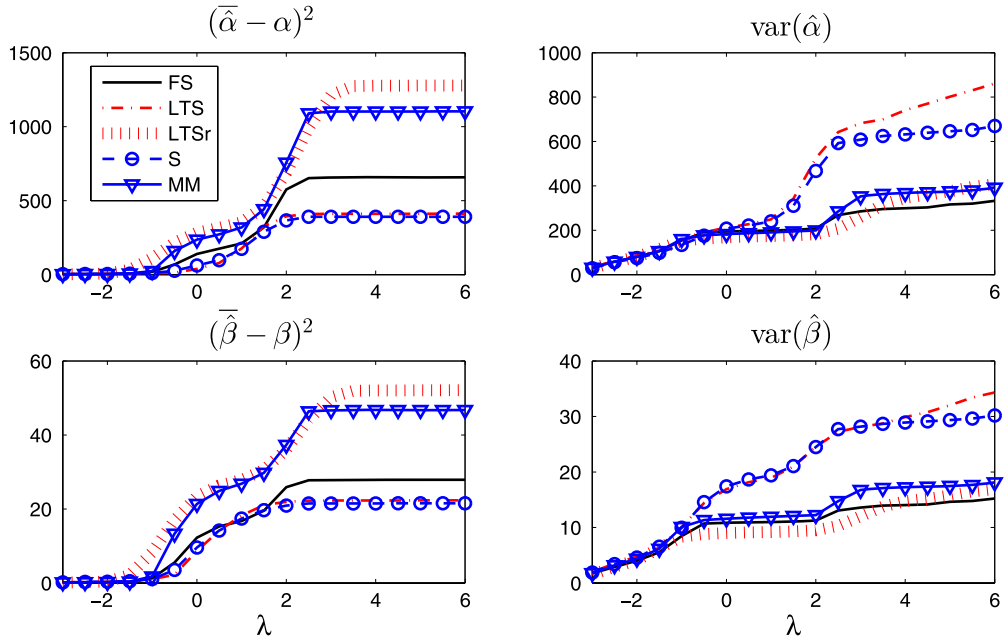}

\caption{Example \protect\ref{ex1}. Partial sums over $\Lambda$ of simulated squared
bias and variance of the
five estimators. Left-hand panels squared bias,
right-hand panels variance. Top line $\hat\alpha$, bottom line
$\hat\beta$.}\label{sim1biasvar}
\end{figure*}

Figure~\ref{sim1biasvar} provides a powerful summary of the results on
the variance and bias of the estimates of $\alpha$ and $\beta$ as
$\lambda$ varies. The left-hand panels show the
partial sums of the squared bias over $\Lambda$ and the right-hand
panels show the partial sums of the variances. The values for $\alpha$
are in the top row and those for $\beta$ in the bottom row.

The plots illustrate the trade-off between bias and variance for some
of the estimators. For values of $\lambda$ up to three or so, LTS and S
have the highest variances and the lowest biases and have very similar
properties. Over the same range LTSr and MM have high biases and low
variances. The effect of the modification of LTS to LTSr and S to MM
has, in general, been to reduce variance at the cost of an increase in
bias. The bias values for FS are in between those of these two groups,
but closer to the lower pair of values, especially for estimation of
$\beta$. The variance of FS is close, and ultimately less than, the low
values for LTSr and MM.

The bottom right panel of Figure~\ref{Sim1boxsca} shows that for
$\lambda= 3$, the outliers are becoming distinct from $y_1$. As
$\lambda$ increases further, the two groups become increasingly
distinct, an effect that is evident in Figure~\ref{sim1biasvar}. For
the extreme values of $\lambda$, the horizontal value of the
summed squared bias for all estimators shows that the bias is zero. The
two populations are sufficiently far apart that the asymptotics
defining high breakdown apply. This is achieved for slightly less
separation by MM than LTSr. The plots of partial sums of variances, on
the other hand,
increase steadily, since the estimators are always subject to the
effect of the random variability in the observations. The sums of
variances for S and, particularly, LTS are, however, increasing more
rapidly at the ends of the region than those for the other three
methods, a result in line with the rows of boxplots for $\lambda= \pm
3$ in Figure~\ref{Sim1boxsca}.

These plots illustrate the differing performance of the five
estimators. Since this is a paper about robust statistics, we also
looked at plots in which the variance of the estimators was replaced by
the average median absolute deviation from the median. These plots were
close to those of the variances shown here.

\begin{figure*}[b]
\vspace*{6pt}
\includegraphics{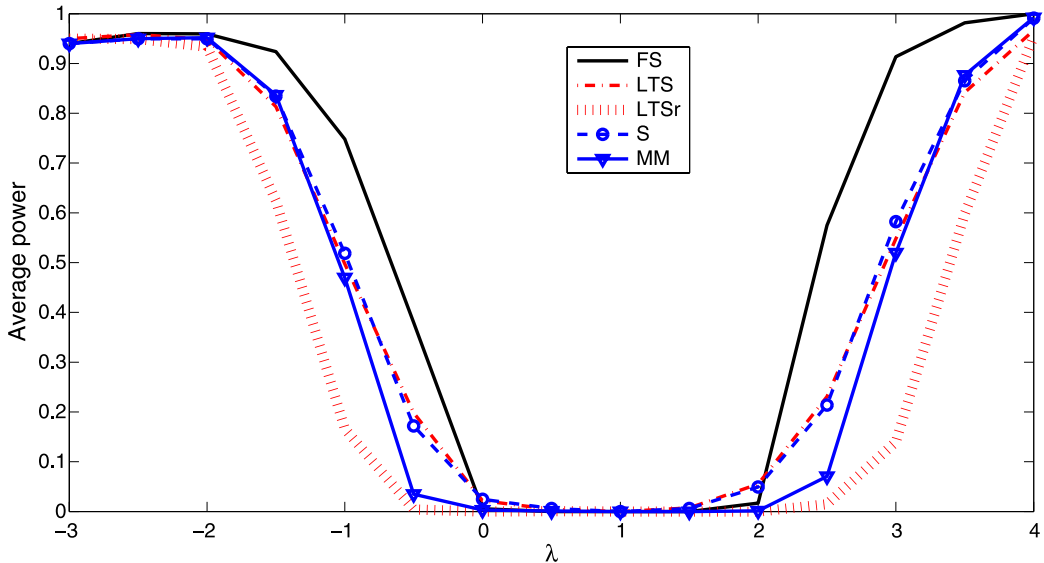}

\caption{Example \protect\ref{ex1}. Simulated average power of the five
procedures over $\Lambda$.} \label{sim1avpow}
\end{figure*}

In addition to good parameter estimates, we would also like our
estimate to signal the presence of outliers if the model fitted to
the data is incorrect. Accordingly, we
calculated the average power, that is, the average number of observations
correctly detected as being contaminated, which is the average number
of detected observations from $M_2$. In testing for the presence of
outliers, we used a test of Bonferronised size $\alpha^*$. The results
are in Figure~\ref{sim1avpow}. Outliers are not detected for central
values of $\lambda$, as the parameter estimates are sufficiently
corrupted by observations from $M_2$ that no observations appear
outlying. As the means of the two populations move apart, the number of
outliers detected increases. Over most of the range FS has the highest
power and LTSr the lowest. The other three estimates lie between these
extremes, with MM having lower power for values of $\lambda$ near zero.
As with any power curves calculated for tests whose exact sizes are not
known, we need to calibrate these findings against the size of the
tests (see Section~\ref{powersec}).\looseness=1

\begin{figure*}

\includegraphics{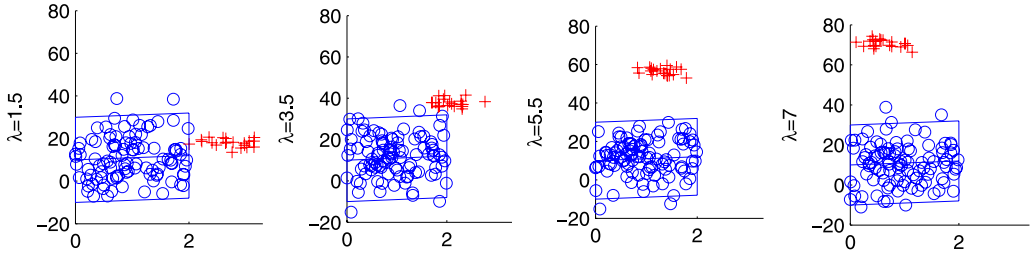}

\caption{Example \protect\ref{ex2}. Simulated data sets with $n_1 =100$ and $n_2 = 20$
for four values of $\lambda$. As $\lambda$ increases, observations from
$M_2$ become close to those from $M_1$ and then become remote again.
The parallelogram defines the region for the empirical overlapping
index.}\label{Sim2scatter}
\end{figure*}

\begin{example}\label{ex2}
In the interests of space we present only a part of
our results, leaving the remainder for the online supplement.
\end{example}

We stay with a single explanatory variable but now choose a trajectory
for $\lambda$ such that $\theta_2^0 \ne\theta_1$, so that most of the
observations $y_2$ are outlying. The parameter values for population 1
were $a=0$, $b=2$, $\alpha=10$, $\beta=1$ and $\sigma_\varepsilon=10$. For population 2, $\Sigma=\operatorname{diag}(4,
0.1), \mu_2 = 3.4$ and $ d = 2$, so that the centres no longer
coincided. Also, $n_2 = 20$. Figure~\ref{Sim2scatter} shows
scatterplots of typical samples for four values of $\lambda$. In the
first, for $\lambda= 1.5$, there is a set of horizontal outliers,
which can be expected not appreciably to affect the estimate of slope.
As $\lambda$ increases, the observations from $M_2$ rise above those
from $M_1$, generating increasingly remote vertical outliers.

\begin{figure*}[b]

\includegraphics{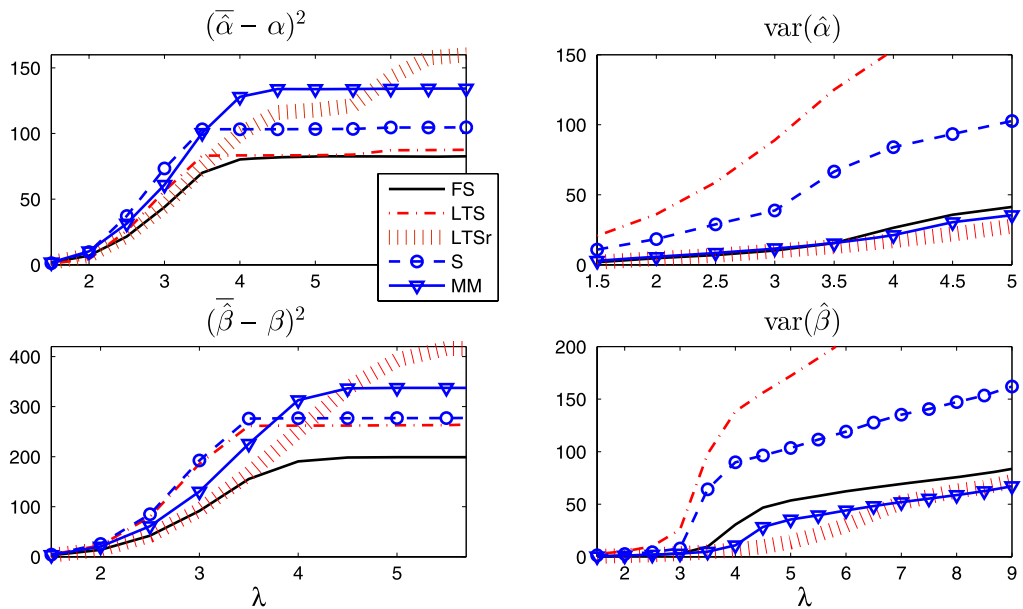}

\caption{Example \protect\ref{ex2}. Partial sums over $\Lambda$ of simulated squared
bias and variance of the
five estimators. Left-hand panels squared bias,
right-hand panels variance. Top line $\hat\alpha$, bottom line
$\hat\beta$.}\label{sim2biasvar}
\end{figure*}

The behaviour of the five estimators for this new situation is
summarised in the partial sum plots of Figure~\ref{sim2biasvar}. The
plots of variances are simply interpreted: S and LTS have high variance
for both $\alpha$ and $\beta$ over the whole range of $\lambda$, with
MM and LTSr having low values which are slightly less than that of FS.

The comparison of biases is less straightforward. The scatterplots of
Figure~\ref{Sim2scatter} suggest that the two populations should be
adequately separated by the time $\lambda= 4$. For lower values of
$\lambda$, S and LTS have similar higher biases for $\beta$. The biases
for $\alpha$ do not show much difference for lower values of $\lambda$.
In the right-hand halves of the plots in Figure~\ref{sim2biasvar}, with
$\lambda> 4$, the two populations are more separated. The plots of
bias show that S and LTS provide unbiased estimates (horizontal plots)
for smaller values of $\lambda$ than does MM. The LTSr estimates are
not unbiased, even for the largest values of $\lambda$. The FS has
excellent properties; it has the lowest bias for both parameters and a
variance which is close to those from MM and LTSr.

The plot of average power for this example in Figure~19 of \citet
{r+a+p:2013b} leads to similar conclusions to those for Example~\ref{ex1} in
Figure~\ref{sim1avpow}. FS has the highest power and LTSr the lowest,
but now the difference between FS and the other rules is much greater.
S and MM have indistinguishable performances, with LTS closer to that
of LTSr.\looseness=1

\begin{example}\label{ex3}
The third example had five explanatory variables ($p = 6$),
independently uniformly distributed on ($0, 2\sqrt10$) with regression
parameters $\beta= 5$ for all variables, $\sigma_\varepsilon=10$ and $n_1 = 200$. For
population 2, $\Sigma= \operatorname{diag}( 100, I_5), \mu_2 = 3, d = 2$ and
$n_2 = 60$.
\end{example}

\begin{figure*}

\includegraphics{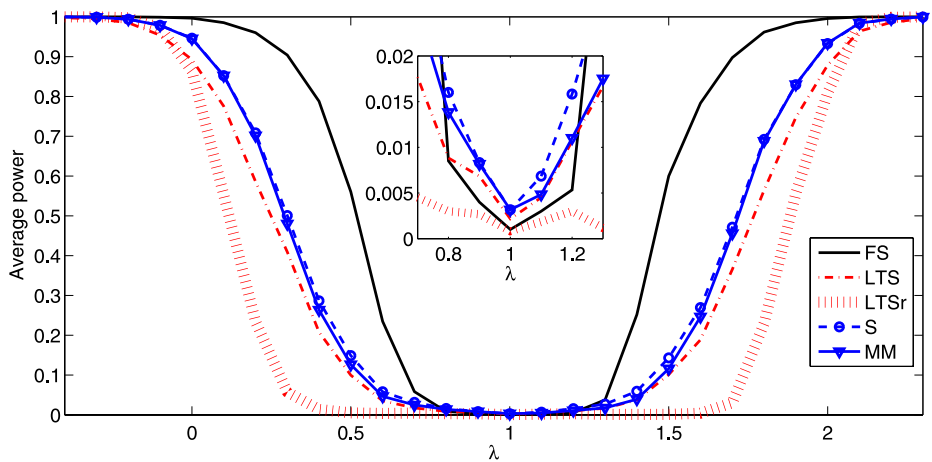}

\caption{Example \protect\ref{ex3}. Simulated average power of the five
procedures over $\Lambda$ with an inset zoom of the central part of the
figure.} \label{sim3avpow}
\end{figure*}

This is a larger example, with $n_1 = 200$ and $n_2=60$. As $\lambda$
increases from $-1$ to 2.6, the outliers ``rise through'' the central
observations. Since $d \ne0$, the centres of the two distributions are
never identical. Unlike our other two examples, this one does not
include outliers at leverage points, so that the differences in
behaviour of the methods are, to some extent, reduced.

With five explanatory variables the major contribution to the mean
squared error of the parameter estimates comes from $\beta$, so we only
consider these values, which are plotted in Figure~21 of \citet
{r+a+p:2013b}. With independent $x_j$, the bias and variance are the
sums of those for the individual components. LTS behaves surprisingly
poorly, with the uniformly highest bias and variance. LTSr and S have
medium behaviour for both properties, with the order reversed for bias
and variance, while MM and FS have the same, lowest values for bias and
similar values for variance until $\lambda= 1$ when that for FS
increases, although staying below that for S. Unlike the other two
examples, the relative behaviour of the estimators is little affected
by the value of $\lambda$, a reflection of the stability of the outlier
pattern over $\Lambda$. Of course, the magnitude of the outliers is
largest for extreme values, but leverage points are not introduced or removed.

The plot of average power is in Figure~\ref{sim3avpow}. As in the other
plots of average power, FS has the highest power and LTSr the least.
The other three estimators have very similar properties to each other.
However, in assessing power we need to be sure that we are comparing
tests with similar sizes. The zoom in the centre of the plot for values
of $\lambda$ close to one shows that we are not, with FS and LTSr,
having the smallest values. For accurate comparisons we need to scale
the other three tests downwards, which will reduce the curves below the
plotted values. However, even when $\lambda= 1$, outliers are still
present and, since $d \ne0$, we are not looking at the null
distribution of the test statistics. We consider null distributions and
the resulting size of tests in Section~\ref{powersec}.

\section{The Numerical Effect of Overlap: Point Contamination}
\label{pointsec}

Point contamination plays an important role in the theory of robust
estimation, for example, in finding conditions of maximum bias in
regression (\cite{rdmartin+:1989}; \cite{berren+:2001}).
Accordingly, we extend our simulations to such contamination. Although
it is a special case of \eqref{bvM2} as $\Sigma\rightarrow0$, there
are new features.

\begin{figure*}

\includegraphics{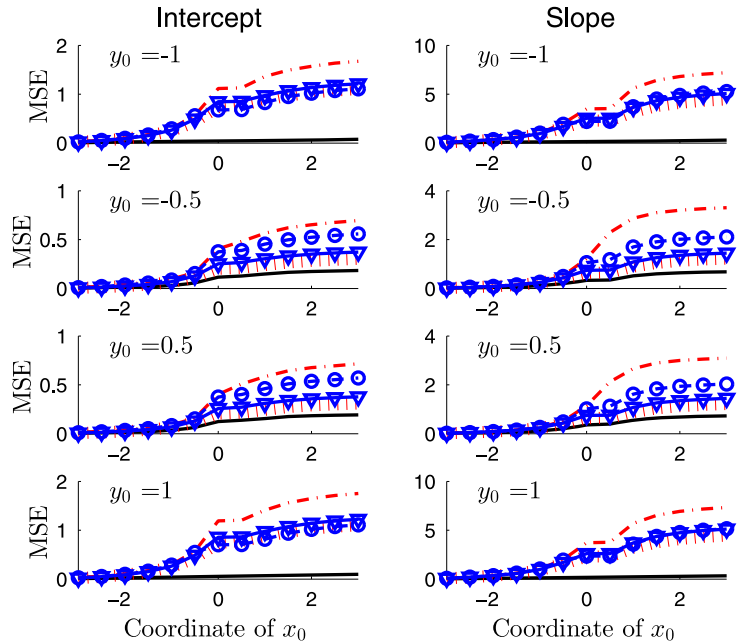}

\caption{Point contamination at $(x_0, y_0)$. Partial sums of mean
squared errors of estimates of $\alpha$ and $\beta$ for five values of
$y_0$ as $x_0$ varies from $-3$ to 3.}\label{sim4horiz}
\end{figure*}

The first feature is the response of the FS algorithm to several
identical observations. As the search progresses, observations are not
only added to the subset used in estimation, but remote observations
are deleted. If several of the identical observations are included,
those outside the point contamination will seem remote and the search
will collapse, since the fitted model will be singular. If such
singularity occurs, we identify all identical observations and force
them to enter at the end of the search. As the figures show, in some
cases this has a powerful beneficial effect on the estimates. The
second feature is that the overlapping index now has the value of
either zero or one.

We took 100 $x$ values between 0 and 1, with the normally distributed
values of $y$ such that approximately 95\% lay between $-0.5$ and 0.5. We
add 30 identical contaminating observations at $(x_0,y_0)$ where both
the vertical and horizontal directions of contamination range from $-3$
to 3.

Figure~\ref{sim4horiz} shows plots of the partial sums of the mean
squared error of the estimates of the intercept and slope for four
values of $y_0$ over a fine grid of values of $x_0$ from $-3$ to 3. The
most notable features are the poor performance of LTS and the good
performance of FS. This is particularly striking in the more extreme
vertical contaminations, $y_0 = \pm1$, where the FS estimates are
virtually unaffected by the thirty outliers.

\begin{figure*}

\includegraphics{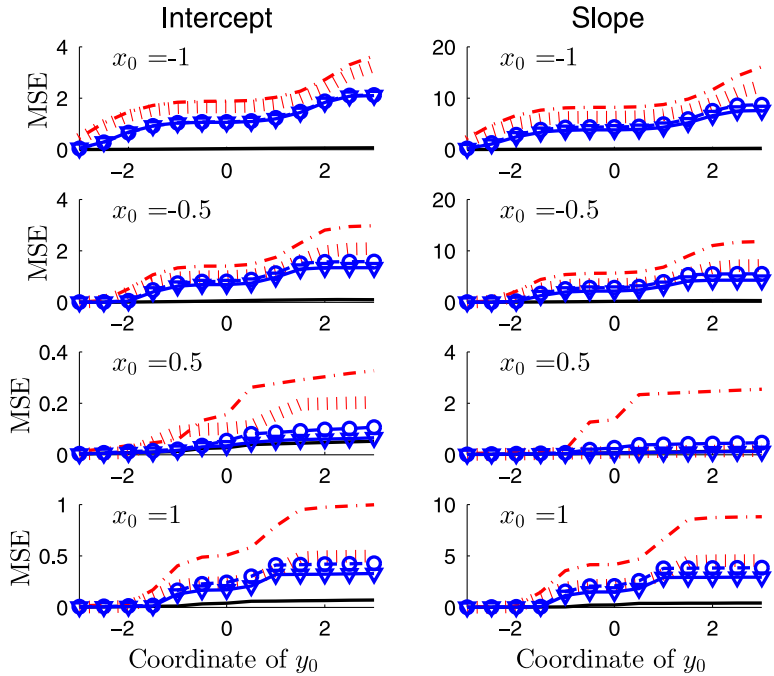}

\caption{Point contamination at $(x_0, y_0)$. Partial sums of mean
squared errors of estimates of $\alpha$ and $\beta$ for five values of
$x_0$ as $y_0$ varies from $-3$ to 3.} \label{sim4vert}
\end{figure*}

In Figure~\ref{sim4vert} we look at the same quantities, as $y_0$
varies for four fixed values of $x_0$: $-1$, $-0.5$, 0.5 and~1. Recall that
the values of $x$ range from 0 to 1, so these plots are not symmetrical
around $x_0 = 0$. The most striking feature is the excellent
performance of the FS, which is by far the best except when the
contamination passes through the centre of ${\cal X}$; even then it is
slightly better than MM and S. LTS behaves particularly poorly when
$x_0 \in{\cal X}$, but is uniformly poorest. S and MM are similar, and
slightly better than LTSr.

The use of point contamination allows sharp comparison of the
algorithms for very robust regression. In the more diffuse situations
of Section~\ref{olapsec} the plots of the power curves, such as those
of Figure~\ref{sim3avpow}, help to strengthen the comparisons.

With this two-dimensional model for contamination it is possible to
explore the properties of the estimators over a grid of values for
$(x_0, y_0)$. With higher dimensional problems, such as Example \ref{ex3}, we
will again need to construct a trajectory $\Lambda$ along which the
point contamination moves.

\section{Size Comparisons}
\label{powersec}

\begin{figure*}[b]

\includegraphics{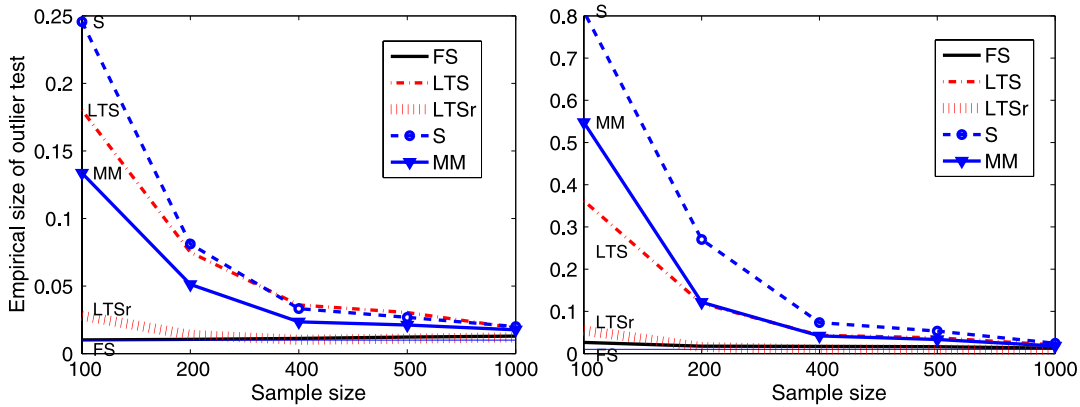}

\caption{Size of nominally 1\% Bonferronised outlier tests for,
left-hand panel, $p = 6$ and for $p = 11$. Note the
different vertical scales in the two panels.}  \label{sizep611}
\end{figure*}

In order to establish the size of the outlier tests, we ran simulations
for sample sizes $n$ from 100 to 1000 for several different dimensions
of problems. The results for $p=6$ and 11 are in Figure~\ref{sizep611}.
In the simulations the samples were allowed to grow with $n$, so that
samples for larger values of $n$ contained those for smaller, leading
to smoother curves. Both the response and the explanatory variables
were simulated from independent standard normal distributions, with all
regression coefficients set to one. Since all methods are affine
equivariant, these arbitrary choices do not affect the results. For
each value of $n$ we present the average of 10,000 simulations, in
which we counted the number of samples declared as containing at least
one outlier, with the tests conducted at the 1\% Bonferronised level.

The figure shows that, for three out of the five rules, the sizes are
very far from the nominal value of 1\%. For $n =100$ the sizes for MM,
LTS and S when $p = 6$ range between 0.13 and 0.25. For $p = 11$ the
range for these rules is 0.36 to 0.81. The sizes decrease with $n$, but
are even so still around 2\% for these rules when $n = 1000$. The size
for LTSr is closer to nominal, being around 3\% and 6\% for $n = 100$
and decreasing rapidly with $n$. Only FS has a size around 1\% for both
values of $p$ and all~$n$.\looseness=1

These calculations of size show that FS is correctly ordered as having
highest power. The curves, such as those in Figure~\ref{sim3avpow}, for
LTSr do not need appreciable adjustment for size. However, size
adjustment for MM, LTS and S may well lead to procedures with less
power than LTSr.

A simple method of adjusting power for size is a normal, or logistic,
plot of the power curves, as in Figure~8.12 of \citet{aca:85}, when the
slope of the curve indicates power and the intercept size. Although
such a comparison would be possible here, our purpose is not to
establish the exact properties of outlier tests. Rather we are
concerned with introducing a general framework for the comparison of
methods for very robust regression.

\begin{figure*}

\includegraphics{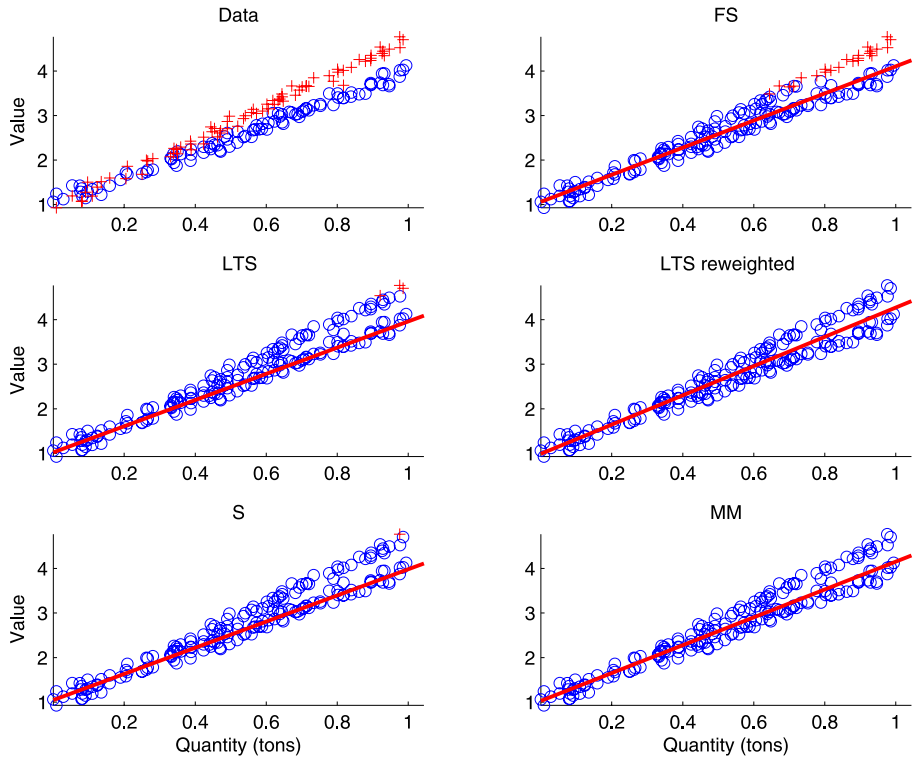}

\caption{Trade data: results of five very robust analyses. Reading
across: $\circ$ simulated regression data and outliers $+$; fitted FS
line and outliers $+$; LTS, reweighted LTS, S and MM estimators.} \label{tradefit}
\end{figure*}

\section{Trade Data Again}
\label{tdsec2}

We began our discussion of very robust regression in Section~\ref{tdsec} with the trade data plotted in Figure~\ref{TDsca}. We now
conclude with a plot of the fitted lines and of the outliers identified
by the five methods we have been comparing. The results are in
Figure~\ref{tradefit} where the top left-hand panel repeats the plot of
the data. The other panels show that FS, LTS and S all provide fits to
the lower of the two lines evident for the higher values of value and
quantity. The other two methods, reweighted LTS and MM, provide fitted
lines which lie more between the groups. Only FS indicates that there
are a large number of outliers which might perhaps be modelled
separately. These results are in line with the conclusions to be
expected from the simulation results of earlier sections, particularly
the low power of the outlier tests for all except FS. However, the
power comparisons combined with the size calculations of Section~\ref{powersec} show that we cannot change the level of the tests without
damaging the size of the test when there no outliers and so identifying
far too may outliers in the null case.

\section{Discussion}
\label{discusssec}

The largest contrast between estimators is shown in the figures for
point contamination of Section~\ref{pointsec}. The relatively poor
behaviour of LTS recalls the impression of \citeauthor{ch:90} [(\citeyear{ch:90}), Section~6],
that the related MVD method for multivariate data finds ``Outliers
Everywhere.'' The superior performance of FS comes from the
data-dependent flexibility of the number of observations included in
the final fit.

Several authors, for example, \citet{chw:93} and \citet{hawk+o:2002},
have commented on the persistence of the effects of the initial
estimator, even asymptotically. The FS escapes such persistence
because, although the subset used in fitting grows in size,
observations can be deleted as well as added. This provides the
algorithmic flexibility that leads to such good performance in Section~\ref{pointsec}. In addition, the flexibility of the FS combined with
the plotting of diagnostic measures makes possible the detection of
subpopulations in the data, not just the point contamination of Section~\ref{pointsec}. An example of cluster detection is shown in Figure~10
of \citet{a+r:2007}.

There is also some theoretical explanation for the relative behaviour
of the other estimators. In particular, the MM estimator is intended to
improve the efficiency of the S estimator and, indeed, this estimator
has a lower variance in Examples \ref{ex1} and \ref{ex2}. But this is achieved at the
cost of having higher bias than the S estimator. The same is true for
the comparison of LTSr and LTS. For those values of ($x_0, y_0$) in
Section~\ref{pointsec} for which $x_0 \in{\cal X}$, so that there are
no leverage points to introduce serious biases, LTSr and MM are,
respectively, an improvement on LTS and S.

We have illustrated the use of our framework for comparing FS with
methods designed to have a breakdown of 50\%. Of course, the framework
can be used for comparisons with breakdown levels more likely to be
used in practice, such as 20\% or 30\%. The properties of FS, since they
do not depend on a specified breakdown level, will not be changed.

\begin{appendix}
\section*{Appendix: The Theoretical Overlapping Index}\label{appA}

\setcounter{equation}{0}

The response and the explanatory variables lie in a space of dimension
$p+1$. Let these variables be $w$. Then the regression plane can be
written as $b^{T}w - c = 0$. The equation of the normal to the plane
through a point $w_0$ on the plane is
%
%
\begin{equation}
z_1 = w_0 +bd, \label{defnorm}
\end{equation}
where the scalar $d$ is the distance from the plane. The outlying
observations, including the response, have a multivariate normal
distribution. Let these be $W \sim{\cal N}(\mu,\Sigma)$. We require
the probability that $W$ lies on one side of the plane. To obtain this,
rotate $W$ to a set of variables $Z$ with $z_1$ \eqref{defnorm} the
normal to the plane. Integrating out the other $p$ variables shows that
the required probability comes from the marginal distribution of $Z_1
\sim {\cal N}(b^T\mu,b^T\Sigma b)$. Let the distance in the $z_1$
direction from $\mu$ to the plane be $d(c)$. Then, from \eqref
{defnorm}, at the plane $b^Tw = c = b^T\mu+ b^Tbd(c)$, so that
%
%
\begin{equation}
d(c) = \bigl(c - b^T\mu\bigr)/b^Tb. \label{defdc}
\end{equation}
Since the distance $d(c)$ in the $z_1$ direction has been rescaled by
the factor $1/b^Tb$, the required probability is
%
%
\begin{eqnarray}\label{probdc}
&&\operatorname{Pr} \bigl(b^TW > c\bigr) \nonumber\\
&&\quad= \operatorname{Pr}
\bigl(Z_1 > c - b^T\mu\bigr)
\\
&&\quad= \Phi\bigl\{
d(c)b^Tb/\bigl(b^T\Sigma b\bigr)^{0.5}\bigr\} =
\Psi(c) \quad\mbox{say},\nonumber
\end{eqnarray}
where $\Phi$ is the c.d.f. of the (univariate) standard normal
distribution. We require this probability in terms of the regression
model, which we now write as $y = \alpha+ \beta^T x$. Then
\[
b^T = \bigl(1 -\beta^T\bigr),\quad w^T = \bigl(y
x^T\bigr) \quad\mbox{and}\quad c = \alpha.
\]
Finally, we require the probability that $W$ lies between two planes.
For any $x$ the required strip around this model is $y \pm2\sigma
_{\varepsilon}$. The two planes then are defined by constants $c^+ =
\alpha+ 2\sigma_{\varepsilon}$ and $c^- = \alpha- 2\sigma_{\varepsilon}$.
From \eqref{probdc} the required probability is $\Psi(c^+) - \Psi(c^-)$.
\end{appendix}

\section*{Acknowledgements}
The work on this paper was jointly supported by the project MIUR PRIN
\emph{MISURA---Multivariate models for risk assessment}, by the
JRC Institutional Work Programme 2007--2013 of the SITAFS Research
Action, and by the OLAF-JRC project \emph{Automated Monitoring
Tool on External Trade}. Much of it was completed at the Isaac Newton
Institute for Mathematical Sciences
in Cambridge, England, during the 2011 programme on the Design and
Analysis of Experiments.

We are most grateful to the referees whose thoughtful and detailed
comments led to improvement and clarification of our paper.

\begin{supplement}[id=suppA]
\stitle{Supplement to ``{A}
Parametric Framework for the Comparison of Methods of Very Robust
Regression''}
\slink[doi]{10.1214/13-STS437SUPP} 
\sdatatype{.pdf}
\sfilename{sts437\_supp.pdf}
\sdescription{\citet{r+a+p:2013b} includes further analyses
of data. The first is a second motivating example; the other two are
expanded versions of our analyses of Examples \ref{ex2} and \ref{ex3} in the paper.
This material is also available at
\url{http://www.riani.it/pub/RAP13supp.html}, together with further, dynamic
graphics and links to the programs used to generate the results in our paper.}
\end{supplement}


\end{document}